# Operational-dependent wind turbine wake impact on surface momentum flux revealed by snow-powered flow imaging

Aliza Abraham[a,b] and Jiarong Hong[a,b,1]

[a]St. Anthony Falls Laboratory, College of Science and Engineering, University of Minnesota, 2 Third Avenue SE, Minneapolis, MN 55414
[b]Department of Mechanical Engineering, College of Science and Engineering, University of Minnesota, 111 Church Street SE, Minneapolis, MN, 55455
[1]To whom correspondence should be addressed. E-mail: jhong@umn.edu; Phone: 612-626-4562 (ME), 612-624-5102 (SAFL)

**Abstract**
As wind energy continues to expand, increased interaction between wind farms and their surroundings can be expected. Using natural snowfall to visualize the air flow in the wake of a utility-scale wind turbine at unprecedented spatio-temporal resolution, we observe intermittent periods of strong interaction between the wake and the ground surface and quantify the momentum flux during these periods. Significantly, we identify two turbine operational-dependent pathways that lead to these periods of increased wake-ground interaction. Data from a nearby meteorological tower provides further insights into the strength and persistence of the enhanced flux for each pathway under different atmospheric conditions. These pathways allow us to resolve discrepancies between previous conflicting studies on the impact of wind turbines on surface fluxes. Furthermore, we use our results to generate a map of the potential impact of wind farms on surface momentum flux throughout the Continental United States, providing a valuable resource for wind farm siting decisions. These findings have implications for agriculture in particular, as crop growth is significantly affected by surface fluxes.

As a renewable and affordable form of energy, wind power is growing rapidly, with a tenfold increase in demand and deployment expected by the year 2050 (Veers et al. 2019). With such expansion will come increased interaction between wind turbines and their surroundings, particularly in the area of agriculture, as the richest land-based wind resources in the United States overlap with land used for growing most of the country's wheat and corn (Rajewski et al. 2013). Wind turbines leave air with lower energy and velocity but increased turbulence in their wake, influencing the flux of momentum, heat, carbon dioxide, and water vapor between the ground surface and the air. Changes in these fluxes affect the air temperature near the surface, which in turn has implications for plant growth in the surrounding area (Rajewski et al. 2013, Li et al. 2018, Xu et al. 2019), though these effects are still not fully understood. Furthermore, prior investigations into the impacts of wind farms on surface temperatures have presented inconsistent results, with several reporting warming during the day (Zhou et al. 2012, Li et al. 2018, Miller & Keith 2018, Xia et al. 2019), others recording cooling during the day (Baidya Roy & Traiteur 2010, Rajewski et al. 2013), and still others observing no significant change in daytime temperature (Baidya Roy et al. 2004, Smith et al. 2013, Archer et al. 2019). These conflicting reports show a lack of fundamental understanding of the mechanism involved in the interaction between the wind turbine wake and the ground surface, caused by limitations in the techniques used to study the phenomenon. Laboratory-scale experiments and computational simulations do not capture the full complexity of utility-scale wind turbine behavior, including dynamic turbine operation and stochastic atmospheric conditions. On the other hand, conventional field-scale flow diagnostic techniques lack the resolution required to capture the instantaneous flow field and coherent turbulent structures involved in the interaction, needed to probe the underlying mechanism.

**Visualization of highly intermittent wake-ground interaction**
To overcome the limitations of previous studies, we implement flow visualization using natural snowfall to investigate the interaction between a utility-scale wind turbine wake and the ground surface at high spatial and temporal resolution. Using a light sheet and camera to capture video of snowflake motion in the wake of a 2.5 MW turbine, we can extract the wind velocity distribution over the illuminated plane (Hong et al. 2014). In this study, the field of view is centered 47 m (0.5 times the rotor diameter, $D$) downstream of the turbine. It intersects the ground and extends up to 49 m, capturing the interaction between the vortices shed from the blade tips and the ground surface (Fig. 1*a*). When these bottom blade tip vortices are produced consistently, they appear as a regularly spaced row of approximately circular dark regions in the images, or voids, where the snow is expelled from the region due to the strong rotation of the air (Fig. 1*b*). However, several periods were observed where strong interaction occurred between the bottom tip vortices and the ground surface, identified by the emergence of a significant amount of chaotic voids above the ground but below the elevation of blade tip vortices (Fig. 1*c*). These voids are caused by the preferential concentration of particles in regions of high strain (Squires & Eaton 1991), indicating the presence large velocity gradients which lead to enhanced momentum flux and mixing. To understand the frequency of occurrence and the causes of these periods of strong interaction, a metric for interaction strength was defined, described in detail in the Methods Section. Using this metric, it became clear that these periods of strong interaction occurred intermittently throughout the duration of the dataset (Fig. 1*d*), begging the question of the cause of such drastic changes in wake behavior.

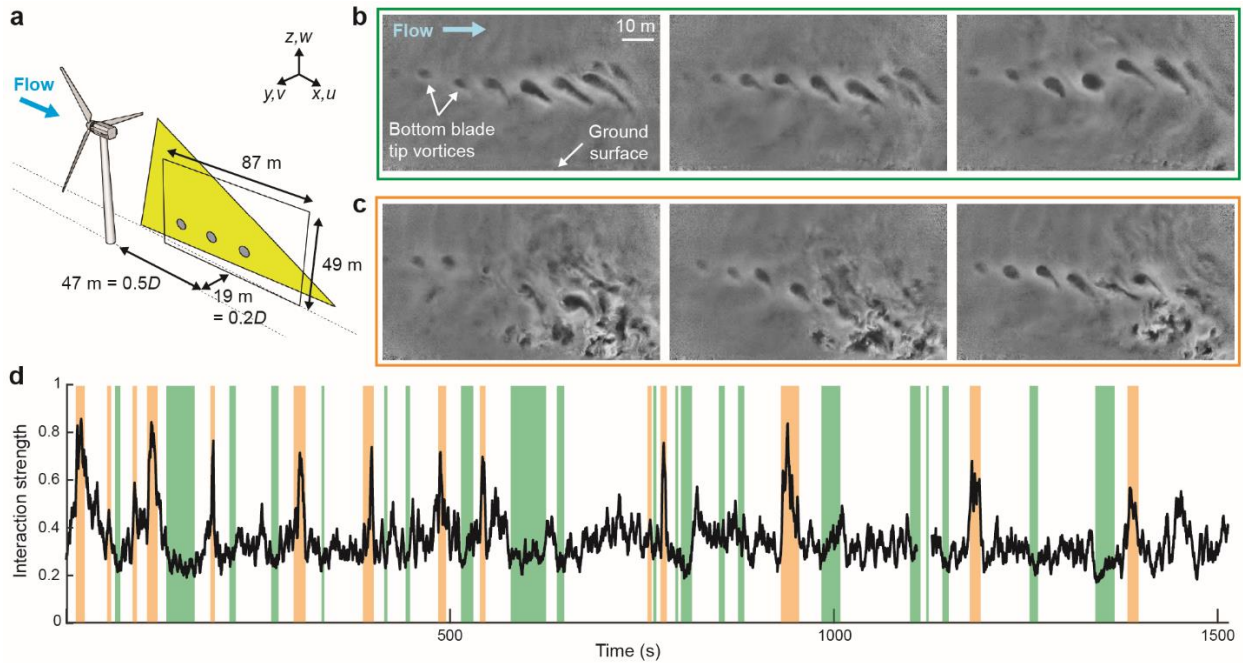

**Figure 1**: Characterization of wake-ground interaction. (*a*) Schematic showing the setup of the experiment, with the yellow triangle indicating the light sheet and the black rectangle showing the field of view. The origin of the coordinate system is at the base of the turbine support tower. (*b*) Sample frames from the video showing periods of consistent bottom blade tip vortex appearance, without visible interaction between the wake and the ground surface. See Video S1 for a sample video clip. (*c*) Sample video frames, identified manually, showing strong interaction between the wake and the surface. See Videos S2 and S3 for sample video clips. (*d*) Automatically characterized interaction strength (black line) compared with manually identified strong and weak interaction periods (orange and green bars, respectively). The orange bars coincide with peaks in the interaction strength, showing the robustness of the automatic characterization method.

**Two turbine operational pathways to strong interaction**
To investigate this question, the interaction strength was compared to a broad range of turbine operational and atmospheric parameters. The coincidence of peaks in interaction strength with peaks in three other parameters was observed: the pitch of the turbine blades (Fig. 2*a*), the change in the tip speed ratio (ratio of the speed of the blade tips to the speed of the wind, Fig. 2*b*), and the power output of the turbine (Fig.

2*c*). However, the overall correlation with each parameter was not very strong, suggesting the periods of strong interaction are caused by a combination of multiple factors. To tease out these relationships, a decision tree was used. Periods of strong interaction were defined as periods with the interaction strength above one standard deviation away from the mean, totaling 7 min of the 55 min of recorded data. Sixteen turbine operational and atmospheric parameters were used to train the decision tree to identify periods of strong interaction. Remarkably, the decision tree identified the same three significant parameters manually detected to have some relationship with the interaction strength (Fig. 2*d*). The decision tree was able to predict the occurrence of strong interactions with 89% accuracy. The events are slightly undercounted, suggesting that there are some periods of strong interaction that occur under conditions not accounted for by the decision tree. However, under the specific operational conditions predicted by decision tree, strong interaction is highly likely to occur. Note that the parameters leading to strong interaction cannot be manually adjusted to induce or mitigate strong interaction with the ground surface, as they are direct responses to atmospheric conditions. They can, however, be used as indicators to predict the occurrence of these periods with reasonable confidence using information already recorded by the turbine controller.

More significantly, the decision tree revealed the existence of two pathways to achieve strong interaction with the ground. The pathway at any given moment is determined by the blade pitch, an indicator of the region of turbine operation (Fig. 2*d*). When the blade pitch < 1.1°, the turbine is operating in region 2 where the atmospheric wind speed is below the rated wind speed and the turbine controller attempts to maximize power production. In this operational region, strong wake-ground interaction occurs when the gradient of the tip speed ratio $\geq 0.09$ s$^{-1}$. The tip speed ratio (ratio between the speed of the blade tip and the incoming wind speed), which is constantly changing in response to the perpetual changes in atmospheric wind speed, determines the spacing between the blade tip vortices. When it increases quickly, the blade tip vortices are pushed closer together, causing them to interact and "leapfrog" over each other. This interaction induces a larger-scale rotation in the flow, pushing the tip vortices closer to the surface and facilitating wake-ground interaction (Fig. 3).

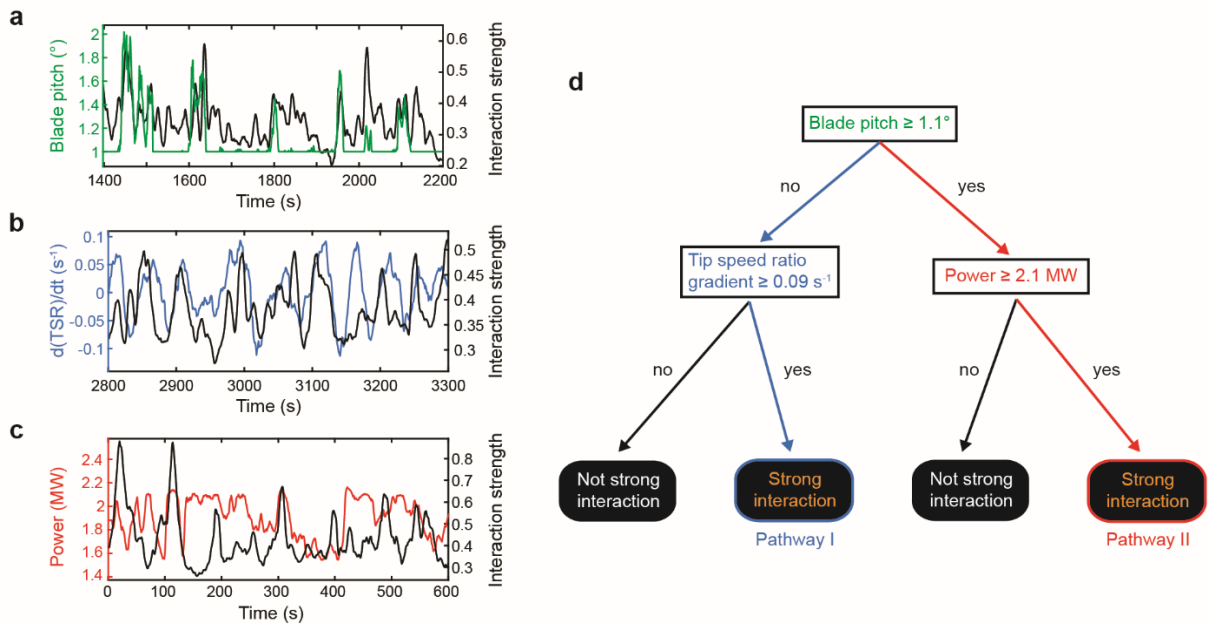

**Figure 2**: Relationship between wake-ground interaction and turbine operational parameters, including (*a*) blade pitch, (*b*) change in tip speed ratio over time, and (*c*) turbine power production. (*d*) Decision tree showing periods of strong interaction can be predicted by the same three parameters identified manually. The two different pathways to strong interaction are shown in blue and red.

The second pathway occurs when the blade pitch ≥ 1.1° and the turbine is operating above the rated wind speed, in region 3 (Fig. 2*d*). In this region, the blade pitch varies to regulate the loading on the turbine structure. This pathway is characterized by power production ≥ 2.1 MW. Two tip vortex features that are characteristic of higher power production are observed in this pathway: wider spread in trajectory angle (Fig. 4*a*) and larger size (Fig. 4*b*) compared to the first pathway. When more energy is extracted from the wind, the velocity deficit in the wake increases, causing the wake expansion to increase to conserve mass, and pushing the tip vortices closer to the ground. Additionally, the size of snow voids associated with tip vortices is correlated with their circulation strength, which increases with turbine power production (Hong et al. 2014). Though these distinctions between the pathways may not be very striking in Figs. 4*a* and *b* due to the dynamic and multi-variate field conditions, a two-sample Kolmogorov-Smirnov test confirms the statistical significance of the differences ($p < 0.001$ for both tip vortex trajectory and size). The velocity and vorticity in the wake confirm these findings, as strong downward flow (Fig. 4*c*) and enhanced vorticity (Fig. 4*d*) are observed around the bottom tip elevation during periods of strong wake-ground interaction in region 3, indicating increased wake expansion and tip vortex strength, respectively.

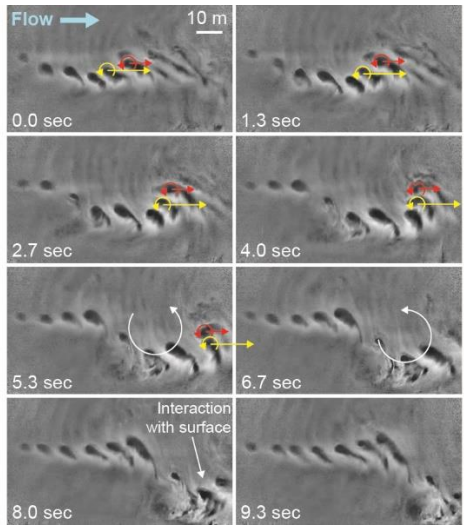

**Figure 3**: Sample image sequence of pathway I to wake-ground interaction, induced by blade tip vortex leapfrogging. One vortex (indicated by a red arrow) moves slower than a vortex behind it (indicated by a yellow arrow), until the rear vortex catches up and the two interaction. The interaction causes a larger-scale rotation (black arrow), which pushes the tip vortices closer to the ground, facilitating interaction between the wake and the surface. See Video S2 for the corresponding video clip.

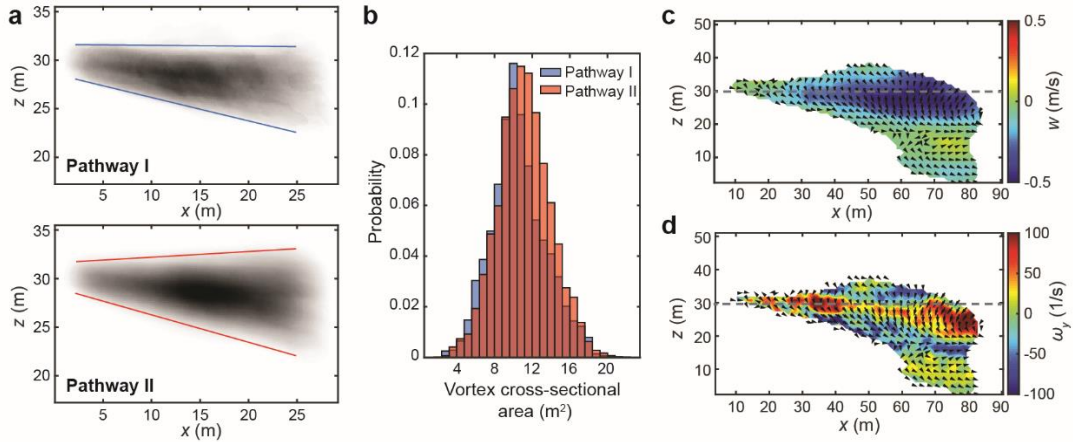

**Figure 4**: Pathway II to wake-ground interaction, caused by (*a*) increased spread in bottom tip vortex trajectory angle (shading indicates the probability of finding a tip vortex in each location) and (*b*) increased bottom tip vortex size (defined as the area of the vortex cross-section in the light sheet) compared with pathway I. The images were calibrated using the method described in Toloui et al. (2014) and Dasari et al. (2019), and the tip vortices were extracted for analysis using the image processing method described in Abraham and Hong (2020). Only tip vortices within 0.3*D* of the turbine were included to preclude distortion caused by interaction with the ground or each other. (*c*) Average wake vertical velocity (*w*) field conditionally sampled for periods of strong interaction in region 3,

subtracted from ensemble averaged region 3 vertical velocity field (original vector fields shown in Fig. S3). Strong downward flow is observed around the bottom blade tip elevation, indicated by a gray dashed line, revealing increased wake expansion during periods of strong interaction in region 3. (*d*) Average wake spanwise vorticity ($\omega_y$) field conditionally sampled for periods of strong interaction in region 3, subtracted from ensemble averaged region 3 vorticity field. Enhanced vorticity is observed around the bottom blade tip elevation (gray dashed line). Video S3 provides a sample video clip of pathway II wake-ground interaction.

**Impact of wake-ground interaction on surface momentum flux**

The abovementioned wake-ground interaction can lead to strong variation of surface momentum flux during the operation of a utility-scale turbine. The vertical profile of the mean momentum flux in the wake compared with that outside the wake is consistent with laboratory scale studies (Chamorro & Porté-Agel 2009, Zhang et al. 2013), showing a strong downward flux just below the bottom tip height and a strong upward flux just above (Fig. 5*a*). On average, no significant increase in downward momentum flux near the surface is observed compared to measurements outside the wake. However, when the average momentum flux is calculated over periods of strong interaction only, a significant increase in downward flux is observed. At an elevation of 10 m, the standard for near-surface atmospheric measurements, the average flux during strong interactions is almost an order of magnitude stronger than that outside of the wake ($-2.1 \times 10^{-3}$ vs. $-0.4 \times 10^{-3}$, nondimensionalized by $U_{\text{hub}}^2$, where $U_{\text{hub}}$ is the mean incoming wind speed at hub height). Though these values are small, they are comparable to normalized surface momentum flux values reported in previous wind tunnel and field studies (e.g., Chamorro & Porté-Agel 2009, Chamorro & Porté-Agel 2010, Zhang et al. 2013, Archer et al. 2019). More significantly, there are some periods of strong interaction where the flux is another order of magnitude stronger (Fig. 5*b*), highlighting the intermittency of the wake effects on surface fluxes. Separating the flux during periods of strong interaction into the two pathways shows that pathway II causes a stronger average flux increase than pathway I by a factor of 1.8 (Fig. 5*c*). Further, pathway II is responsible for the periods of strongest flux. These results show that the extent of wind turbine wake interaction with the ground surface is highly dependent on turbine operating conditions.

Additional data taken from the meteorological tower (met tower) located 170 m (1.8*D*) south of the turbine supplements the flow visualization findings. When the wind blows from the north, the met tower is located within the wake of the turbine and the sonic anemometer at the 10 m elevation can be used to calculate the surface momentum flux. From nearly nine years of stored data, 615 20-min periods are selected where the wind is coming from the north and the turbine is operating in region $\geq 2$. The flux is then conditionally sampled based on the two enhanced flux pathways uncovered from the decision tree. Pathway II enhances flux at the met tower by a factor of 1.8, the same factor calculated from the flow visualization data, while pathway I does not significantly change the surface flux 1.8*D* downstream (Table 1). The different impacts of the two pathways can likely be explained by the different mechanisms for each. Pathway I occurs due to tip vortex leapfrogging in the extreme near wake (<1*D*), as observed in the flow visualization data. Therefore, by the time the wake has travelled further downstream, this effect has dissipated. In the pathway II case, however, the enhanced mixing is caused by the strength of the tip vortices and the wake rather than an isolated event, allowing the effect to persist as the wake travels downstream.

Furthermore, the effect of atmospheric stability on wake-induced momentum flux enhancement is investigated using the more than 200 hours of met tower wake data. The Bulk Richardson Number ($R_B$), a dimensionless number that quantifies the effect of buoyancy due to temperature gradients versus shear-generated turbulence, is used to categorize each 20-min period as stable, neutral, or unstable. The greater the value of $R_B$, the more stable the atmospheric boundary layer, indicating less turbulent mixing is occurring. This categorization reveals that the pathway II surface momentum flux enhancement is strongest when the boundary layer is stable and weakest when it is unstable (Table 1). These findings are consistent with the observations of multiple previous studies that show that the wind turbine wake and the coherent structures within it are stronger and persist longer with increasing atmospheric stability (e.g., Magnusson & Smedman 1994, Chamorro & Porté-Agel 2010, Hansen et al. 2012, Zhang et al. 2013, Abkar & Porté-Agel 2015), providing further support for a stronger wake and tip vortex as the mechanism behind pathway II surface flux enhancement.

**Table 1**: Mean momentum flux for each pathway under different atmospheric stabilities using 205 hours of data from the met tower located 1.8*D* downstream of the turbine.

| $\overline{u'w'}/U_{\text{hub}}^2 \times 10^{-3}$ | Not strong interaction | Strong interaction – Pathway I | Strong interaction – Pathway II |
|---|---|---|---|
| All atmospheric stabilities | -1.1 | -1.1 | -2.0 |
| Stable ($R_B \geq 0.25$) | -0.8 | -0.9 | -2.5 |
| Neutral ($0 \leq R_B < 0.25$) | -1.4 | -1.3 | -1.9 |
| Unstable ($R_B < 0$) | -1.2 | -1.3 | -1.5 |

**Reconciling discrepancies in previous studies**

The dependence of wake-surface interaction on operating conditions, a factor that was not considered in previous studies, allows us to propose an explanation to reconcile the conflicting results from field studies in the literature. (see Supplemental Material for figures supporting the following analysis). Baidya Roy and Traiteur (2010) observed cooling during the day at a California wind farm in the summer of 1989, with the period with the largest temperature change occurring in the afternoon (Fig. S5*a*). Based on 5-min resolution simulated wind speed data, wind speeds are highest in the afternoon during this time of year, suggesting flux enhancement through pathway II was occurring (Fig. S5*b*). This enhanced flux likely caused cooling due to the presence of aquifer recharging ponds located between the wind farm rows (Archer et al. 2019). Zhou et al. (2012) investigated the effect of a Texas wind farm on temperature changes in the winter and summer, finding significant temperature increases at night, particularly in the summer. This observation could be caused by enhanced mixing bringing warmer air down to the surface, which is typically cooler at night. Simulated historical data shows that the mean wind speed at the site of the investigation is higher at night, especially during the summer (Fig. S6*a*). However, the rated wind speed of the turbines in the wind farm is 12 m/s, while the maximum mean wind speed is 8.5 m/s which may not be high enough to push the turbines into region 3 for significant amount of time. On the other hand, the standard deviation of wind speed is higher in the evening, particularly in the summer (Fig. S6*b*). The standard deviation, calculated for each hour using 5-min resolution data, represents the level of wind speed fluctuations that cause changes in tip speed ratio. The occurrence of high wind speed fluctuations in the evening indicates that pathway I flux enhancement could be responsible for the observed temperature increase. Smith et al. (2013) reported an increase in surface temperature at night in a wind farm in the United States Midwest. The largest temperature changes occurred during the periods of highest wind speed (Fig. S7), suggesting flux enhancement through pathway II. Rajewski et al. (2013) compared surface warming to boundary layer stability in an Iowa wind farm. They observed warming when the boundary layer was slightly stable (Fig. S8*a*), which typically occurs in the evening and morning (Kumar et al. 2006). Simulated historical data shows the highest standard deviation of wind speed also occurs during these periods (Fig. S8*b*), indicating the temperature changes were caused by pathway I flux enhancement. Additionally, the rated wind speed for the turbines investigated was 14 m/s, significantly higher than the maximum mean wind speed of 7.2 m/s, so pathway II is less likely to be responsible for the observed changes. Finally, Archer et al. (2019) investigated the impact of a single turbine in Delaware on surface fluxes. They did not observe any flux increases in the turbine wake, potentially due to the fact that the periods investigated had very low wind speeds (~5 m/s at hub height), during which the turbine would be operating below region 2. In our experiments, we did not observe any significant interaction between the wake and the surface when the wind speed was so low, as the tip vortices and wake expansion are not strong enough for the wake to reach the ground (see Video S4). Though we acknowledge that many other factors may influence the previously reported findings, this analysis demonstrates that turbine operational conditions cannot be neglected when evaluating the impact of wind turbines on their surroundings.

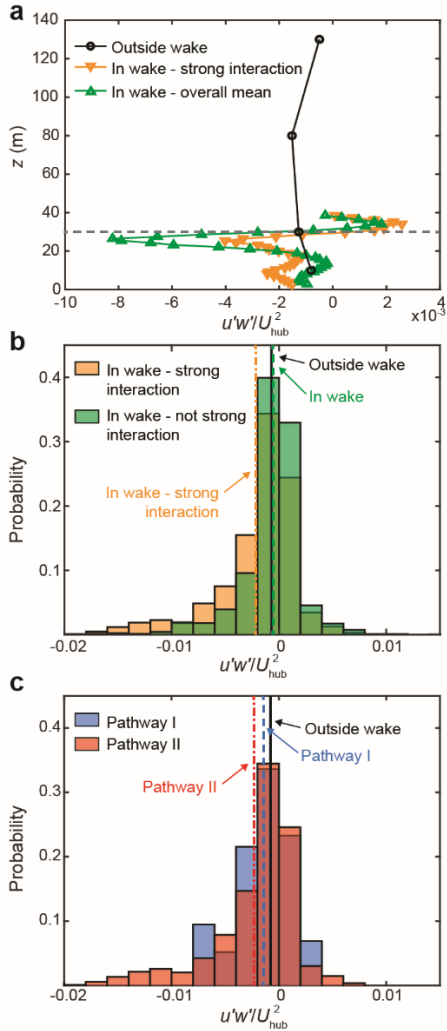

**Figure 5**: The effect of wake-ground interaction on surface momentum flux. (*a*) Profile of the average momentum flux outside of the wake, inside of the wake, and inside the wake only during periods of strong interaction, normalized by the incoming wind speed. The horizontal dashed line indicates the bottom blade tip elevation (30 m). (*b*) Histogram of the normalized momentum flux in the wake at the 10 m elevation for periods of strong interaction and periods not identified as strong interaction. The average of each case, including the momentum flux outside the wake, is indicated by a vertical line. (*c*) Histogram of the normalized momentum flux in the wake at the 10 m elevation for each strong interaction pathway. The average of each case, including the momentum flux outside the wake, is indicated by a vertical line.

**Agricultural implications**

The impact of wind turbines on surface fluxes has significant implications for agriculture. In the United States in particular, much of richest wind resources coincide with regions of high agricultural productivity (Fig. 6*a*). The question of whether wind farms would be beneficial or detrimental to agriculture remains open. The enhanced mixing near the surface caused by wind turbine wakes could be beneficial to plants during the day, as evapotranspiration, the process that cools plants and enables the diffusion of gasses necessary for photosynthesis, is stronger when light is available (Sakurai Ishikawa et al. 2011). On the other hand, enhanced mixing at night when no photosynthesis is occurring could increase evapotranspiration when it is not needed, causing the plants to dry out and embolize (Sperry & Tyree 1988). To quantify the balance of these two effects, a diurnal surface momentum flux impact index, $I_{DSMF}$, is defined as the difference between the wake-induced mixing enhancement at noon and the mixing enhancement at midnight, multiplied by the percentage of the land used for agriculture. Only the summer months (June, July, and August) were included to eliminate the effect of seasonal variability. Overall, the mixing enhancement at night is stronger than that during the day in most of the US, with the exception of parts of Florida, the Gulf Coast, Appalachia, and scattered locations across the Western US (Fig. 6*b*). However, looking at the two pathways for flux enhancement separately provides more nuanced insight. The impact from pathway I is dominant during the day (Fig. 6*c*), while the impact from pathway II is more prevalent at night (Fig. 6*d*). Fortunately, pathway II can be avoided without modifying the turbine control algorithm by ensuring that the turbines placed in the locations in red in Fig. 6*d* have high enough rated wind speeds to operate in region 3 only very rarely. This design choice would potentially mitigate the detrimental effects of enhanced mixing at night, while introducing a higher probability of agriculturally beneficial enhanced daytime mixing. Harnessing the potential positive side effects of turbine wake flux enhancement can ensure future wind farm installations provide benefits to farmers who choose to host them, both in the form of additional income and additional crop yield. The aforementioned findings can be incorporated into crop growth models to more accurately predict the agricultural productivity of fields that are co-located with wind farms.

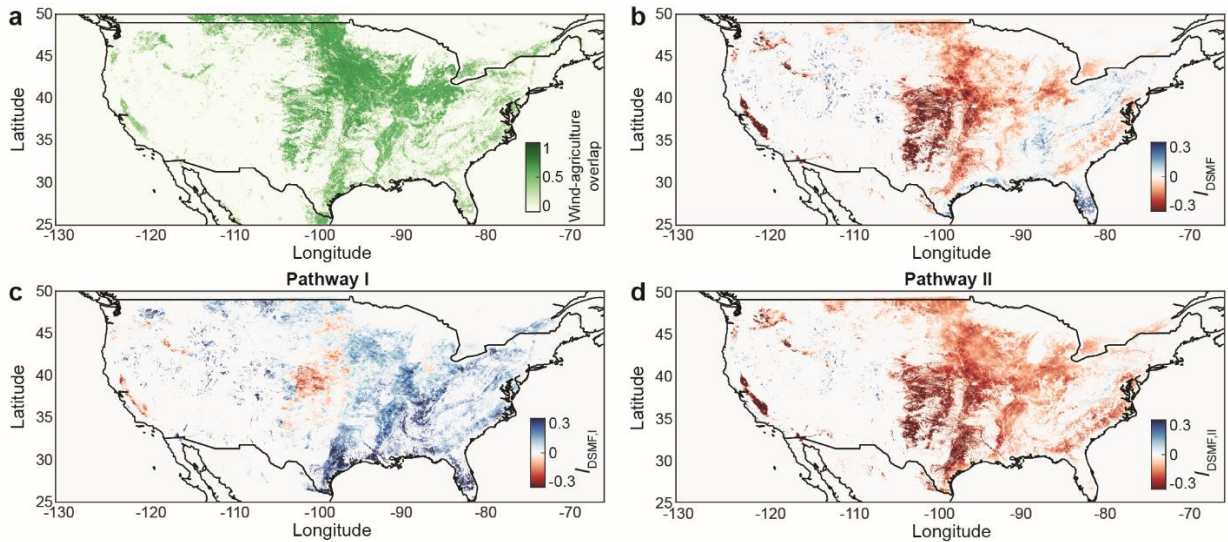

**Figure 6**: Interaction between agricultural land and potential wind farms in the Continental United States. (*a*) Overlap between wind resources and agricultural land. (*b*) Diurnal impact of potential wind farms on surface momentum flux enhancement, considering both pathways, with a positive index indicating more mixing during the day and a negative index indicating more mixing at night. (*c*) Diurnal impact of potential wind farms on mixing, only considering pathway I. (*d*) Impact of potential wind farm on mixing, only considering pathway II.

**Discussion**

The results of the current study can provide insight into the impact of wind turbine wake-ground interaction on other environments in addition to agriculture. Previous studies have shown that wind farms can cause warming of lakes downwind due to increased evaporation (Abbasi & Abbasi 2000) and offshore wind farms can affect the properties of the surrounding ocean waves (Christensen et al. 2013). As another potential benefit of the enhanced mixing observed here, wind farms could be coupled with solar farms, as surface cooling is a key issue limiting solar panel efficiency (Siecker et al. 2017). The results of the current study have global implications; multiple studies have shown that large-scale wind farms can affect global climate, but the magnitude of the impact is not consistent between them (Keith et al. 2004, Li et al. 2018, Miller & Keith 2018). The dependence of wake-ground mixing on incoming flow and turbine operational conditions can be incorporated into the global climate models used to assess wind farm impacts in order to achieve more reliable results.

Though the current study is largely limited to the near wake of a single utility-scale turbine, the physical insights gained are broadly applicable. We observed that enhanced mixing between the wake and the ground happens under turbine operating conditions that are likely to occur regardless of the specific turbine design and turbine layout in a farm. Furthermore, the effects of the near wake are likely significantly stronger than those of the far wake due to the presence of the strong coherent structures shed from the blade tips. In the near wake, the blade tip vortices generate a concentrated region of increased vorticity and turbulence intensity, which weakens significantly after the tip vortices begin to break down (Sherry et al. 2013, Lignarolo et al. 2014, Lignarolo et al. 2015). Beyond the point of tip vortex breakdown, random mixing leads to re-entrainment of momentum and re-energizing of the wake. However, according to the Biot-Savart Law, coherent vortices induce a velocity in the fluid at a distance *r* away, while random turbulent fluctuations do not have this property. The strong coherent vortices (i.e., blade tip vortices) that have the ability to induce a velocity at the ground surface only exist in the near wake until ~2*D*, so the interaction between the wake and the ground is likely to be strongest within this region.

When extending these results to a whole wind farm, the result will likely be spatial and temporal heterogeneity. The near wake of each individual turbine will experience periods of enhanced surface momentum flux, while the far wakes of individual turbines and the wake of the wind farm as a whole will not see the same effects. This observation brings to light the shortcomings of many wind farm parameterizations in climate models. Wind farms are often modelled as roughness elements (Keith et al. 2004, Wang and Prinn 2010, Li et al. 2018) or momentum sinks and turbulent kinetic energy (TKE) sources (Baidya Roy et al. 2004, Fitch 2015, Miller and Keith 2018, Xia et al. 2019), neither of which captures the aforementioned heterogeneity. The actual wind turbine wake, particularly the near wake, is more complicated than a uniform TKE increase and momentum deficit, with coherent structures and a complex wake profile (Abraham et al. 2019). Even higher-fidelity models such as large eddy simulations (LES) often do not account for the stochasticity of the atmosphere or the corresponding changes in turbine operation (e.g., pitch, hub speed) that are responsible for the intermittency of the flux enhancement observed in the current study. The spatial and temporal averages of the wind farm boundary layer mask the local extremes that could significantly affect agriculture and other environments around the wind farm.

However, we urge the reader to use caution when extending our findings, as the physical understanding of our results was derived from intermittent occurrences during an experiment conducted under limited atmospheric conditions at a single site. Under different atmospheric stabilities or surface roughnesses, the tip vortices may interact differently with the ambient turbulence, leading to additional mechanisms for wake-ground interaction. Furthermore, heat and moisture fluxes in the wind turbine wake were not measured in the current study, though they will certainly impact plant growth and may be affected by the wind turbine wake differently than momentum flux. Ambient conditions such as humidity may also change the implications of our findings, as in extremely dry climates, plants may be at risk of drying out during the daytime, while humid climates could provide enough moisture to prevent embolism, even at night. Additionally, we have not considered the beneficial impact of enhanced flux on temperature. For example, enhanced mixing that warms air near the surface at night could prevent crops from freezing in cold climates (Crawford and Leonard 1960). To determine the crop growth response to the enhanced flux in different climates, controlled laboratory studies must be conducted, as a multitude of factors can impact agricultural productivity (e.g., irrigation, soil quality, fertilization, cloud cover, etc.). Finally, the current study uses simulated data to determine the diurnal surface momentum flux impact. Therefore, when deciding whether to build a wind farm in a specific location, the framework provided here can be used with more accurate local wind data to improve the reliability of the predicted effect on the surrounding environment.

**Materials and methods**

**Eolos field site.** The experiment was conducted at the University of Minnesota Eolos field site in Rosemount, MN (Fig. S1). The site consists of a heavily instrumented 2.5 MW Clipper Liberty C96 wind turbine and a 130 m met tower. The turbine is a three-bladed, horizontal-axis, pitch-regulated, variable speed machine with a 96 m rotor diameter mounted atop an 80 m tall support tower. A SCADA system is located at the hub, and strain gages are mounted around the tower base and along the blades. The SCADA system recorded incoming wind speed and direction at a frequency of 1 Hz and hub speed, blade pitch, power generated, and rotor direction at 20 Hz for this study. The met tower is located 170 m south of the turbine and comprises wind speed, direction, temperature, and humidity sensors at six elevations ranging from 7 m to 129 m (details provided in Fig. S1). All the sensors record data 24 hours a day since November 2011, which is stored on database servers.

**Flow visualization using natural snowfall.** Super-large-scale flow visualization using natural snowfall is described in detail in Hong et al. (2014), but a brief summary is provided here. Natural snowflakes serve as the environmentally benign seeding mechanism for a large volume in the near wake of the turbine over several hours. They have strong light-scattering capabilities, and sufficient traceability for large-scale flow structures. Our previous publications have conducted detailed analysis of the traceability of snowflakes for large-scale flow measurements (e.g., Hong et al. 2014, Toloui et al. 2014, Dasari et al. 2019). In the current study, snowflake patterns representing coherent flow structures are tracked rather than individual snowflakes. This specific concept is validated in Dasari et al. (2019).

**Experimental setup.** The flow visualization setup is composed of an optical assembly for illumination and a camera. The optical assembly includes a 5-kW collimated searchlight with a 300 mm beam diameter (divergence < 0.3°) and a curved reflecting mirror to project the horizontal beam into a vertical light sheet. The sheet expansion angle is controlled by adjusting the mirror curvature. The local wind direction was used to align the light sheet parallel to the wind with an angle of 90.4° clockwise from North, 47 m downstream of the turbine and offset 19 m from the plane of the tower in the spanwise direction (Fig. 1*a*). As the wind direction changed throughout the experiment, the degree of misalignment between the light sheet and the rotor changed within the range of -17.3° to 12.0°, with an average of 0.8°. A Sony A7RII camera with a 50 mm f/1.2 lens was used to capture video data at a frame rate of 30 Hz and size of 1080 pixels × 1920 pixels. The camera was placed 120 m away from the light sheet and tilted up 12.4° with respect to the ground. The setup resulted in a field of view of 87 m × 49 m (streamwise × vertical).

**Experimental conditions.** The experiment took place between 23:00 CST on March 5$^{th}$, 2018 and 01:00 CST on March 6$^{th}$, 2018. The atmospheric and incoming wind conditions during this period were recorded using the met tower and SCADA system. The temperature at hub height stayed relatively constant between -3.1°C and -2.8°C. The variation in temperature between the bottom and top of the wake was approximately 0.8°C. The resulting Bulk Richardson Number was 0.12, indicating an approximately neutral atmospheric boundary layer where turbulence is generated mechanically rather than as a result of thermal gradients. The wind speed varied slightly over the course of the experiment, with instantaneous values between 7 m/s and 11 m/s at the hub, allowing the characterization of the wake under turbine operational regions 2-3. The wind direction was primarily easterly, varying between 73° and 103° clockwise from North. Due to the wind direction, the met tower was not influenced by the wake of the turbine during this time period.

**Interaction detection.** Periods of strong interaction between the wake and the ground surface were first detected using manual inspection. When coherent structures were observed between the bottom blade tip vortices and the ground, the corresponding time period was labelled as "strong interaction". When no structures were observed, it was labelled as "no interaction". To make the process more robust and objective, an automatic detection method was developed. First, the images were enhanced using moving average background division with a 1000 frame window, adaptive histogram equalization, and wavelet-based denoising. Next, the region of the images below the bottom blade tip vortices was smoothed using a 9-pixel Gaussian filter and the pixel intensity gradient was calculated (Fig. S2). The mean of the gradient over this region was used to determine the interaction strength, as many strong voids are observed during periods of strong interaction, yielding a large intensity gradient. This metric was found to accurately identify the same periods of strong interaction identified manually, showing it is an effective method for automatic interaction detection (Fig. 1*d*). All image processing was conducted in MATLAB.

**Vector calculation.** To quantify the velocity field, the image distortion induced by the tilt angle of the camera is first corrected following the method described in Toloui et al. (2014). Additionally, the motion of clouds is visible in the background of the video. This is removed by applying a high-pass filter with a cutoff frequency of 0.5 Hz to the video, based on a peak in the video frequency spectrum below the blade pass frequency of the turbine, believed to be caused by the cloud motion. The velocity vectors were calculated using the adaptive multi-pass cross-correlation algorithm from LaVision Davis 8 with an initial interrogation window of 64 × 64 pixels and second pass with interrogation windows of 48 × 48 pixels with 75% overlap. Note that the video does not have sufficient resolution to resolve the individual snow particles in the images, so the snow patterns associated with coherent flow structures provide the signal for the cross-correlation (Dasari et al. 2019). The cross-correlation is applied with a 5-frame skip to ensure sufficient displacement of the structures, resulting in a temporal resolution of 6 Hz. A Hampel filter is applied to the final vector field to remove outliers.

**Met tower data for flux and atmospheric stability calculation.** As the met tower is located south of the turbine at the Eolos site, it is in the wake of the turbine at 1.8$D$ downstream when the wind is blowing from the north. The nearly nine years of recorded data are conditionally sampled for periods of at least 20 min where the wind is between -11° and 12° clockwise from north and the turbine is operating in region $\geq 2$. The 20 Hz 3-component wind data from the sonic anemometer at the 10 m elevation is used to compute the vertical momentum flux, normalized by the incoming hub height wind speed recorded by the SCADA system simultaneously. The met tower flux data is conditionally sampled for atmospheric stability using the Bulk Richardson Number,

$$R_B = \frac{g\Delta\overline{\theta_v}\Delta z}{\overline{\theta_v}[(\Delta\overline{U})^2+(\Delta\overline{V})^2]},$$

where $g$ is gravitational acceleration, $\theta_v$ is the virtual potential temperature, $z$ is the elevation, $U$ is the northerly wind component, and $V$ is the westerly wind component (Stull 1988). The temperature and relative humidity sensors along with the cup and vane anemometers at the 126 m and 7 m elevations were used to perform this calculation for each 20 min time period. Note that, even though the met tower is in the wake of the wind turbine, $R_B$ uses mean values of wind speed and temperature from the top and bottom edges of the wake, which will not be significantly impacted by the wake, as evidenced by the results shown in Figure 5b and as observed in Chamorro & Porté-Agel (2010).

**US wind resources and agricultural land use data.** Agricultural land use data was obtained from the Global Land Cover-SHARE database from the Food and Agriculture Organization of the United Nations (FAO 2013), which provides the percentage of each square kilometer used for cropland. To obtain Fig. 6*a*, this land use data was combined with the mean wind speed at 100 m elevation from the Global Wind Atlas 3.0, a free, web-based application developed, owned, and operated by the Technical University of Denmark (DTU 2019). Five-min resolution simulated wind speed data for the Continental United States during 2014, used to calculate the diurnal surface momentum flux impact index and to compare with previous studies, was obtained from the Wind Integration National Dataset Toolkit (Draxl et al. 2015). Note that these databases were not used to directly calculate the momentum flux. Rather, they are used to predict the likelihood of a wind turbine or farm operating under conditions that would cause enhanced interaction with the surface through either of the two pathways.

**Diurnal surface momentum flux impact index.** The diurnal surface momentum flux impact index ($I_{\text{DSMF}}$) is defined to quantify the net impact of wind turbine wakes on momentum flux enhancement at the ground surface during the day versus the night. To calculate the index, the mean and standard deviation for each hour of wind speed data are first computed. Because the wind speed data has 5-min resolution, 12 data points are used for each calculation. These quantities are then combined into cross-tables with an average mean and standard deviation for each hour of the day in each month (see Fig. S4 for an example). Pathway I for enhanced flux is characterized by changes in tip speed ratio, which is in turn determined by changes in wind speed, while pathway II is characterized by high mean wind speeds that push the turbine into region 3 operation. Therefore, the average standard deviation and mean for the summer months (June, July, and August) at noon and midnight local time are used to calculate the index for pathways 1 and 2, respectively. The index for pathway I is

$$I_{\text{DSMF,I}} = (\text{std}(u)_{\text{noon}} - \text{std}(u)_{\text{midnight}}) * A/\max\left(|(\text{std}(u)_{\text{noon}} - \text{std}(u)_{\text{midnight}}) * A|\right),$$

where $u$ is the wind speed and $A$ is the percentage of land used for cropland. For pathway II, the index is defined as

$$I_{\text{DSMF,II}} = (\text{mean}(u)_{\text{noon}} - \text{mean}(u)_{\text{midnight}}) * A/\max\left(|(\text{mean}(u)_{\text{noon}} - \text{mean}(u)_{\text{midnight}}) * A|\right).$$

The overall diurnal surface momentum flux impact index combines the effects of both pathways:

$$I_{\mathrm{DSMF}} = 0.6 I_{\mathrm{DSMF,I}} + I_{\mathrm{DSMF,II}}.$$

The index for pathway I is weighted with 0.6 because the flux enhancement due to pathway I is 0.6 times that of pathway II.

## Acknowledgements

This work was supported by the National Science Foundation CAREER award (NSF-CBET-1454259), Xcel Energy through the Renewable Development Fund (grant RD4-13) as well as IonE of University of Minnesota. We thank the students and engineers from St Anthony Falls Laboratory, including T Dasari, B Li, Y Wu, S Riley, J Tucker, C Milliren, J Marr, D Christopher, C Ellis for help with the experiments. We also thank Dr. N Davis and Dr. BO Hansen from DTU for assistance with the Global Wind Atlas data.